\documentclass[prd,aps,showpacs,preprintnumbers,amssymb]{revtex4}

\usepackage{color}
\usepackage{epsf}

\def\e3p{$\eta \rightarrow 3 \pi$}

\begin{document}
\title{%
\hfill{\normalsize\vbox{%
\hbox{}
 }}\\
{Dynamical electroweak breaking and contributions to supersymmetry breaking through logarithmic operators}}

\author{Renata Jora
$^{\it \bf a}$~\footnote[1]{Email:
 rjora@theory.nipne.ro}}

\affiliation{$^{\bf \it a}$ National Institute of Physics and Nuclear Engineering PO Box MG-6, Bucharest-Magurele, Romania}

\date{\today}

\begin{abstract}
We introduce in an arbitrary supersymmetric version of the standard model without elementary scalars a logarithmic operator which is irrelevant in the limit in which all symmetries are respected. However through  a dynamical mechanism of the Nambu Jona Lasinio type this operator may produce two fermion vacuum condensates that may account for  electroweak and contribute to supersymmetry breaking. Some of the phenomenological consequences of this set-up are discussed.
\end{abstract}
\maketitle

\section{Introduction}

Beyond  the standard model theories rely mainly on three mechanisms: dynamical symmetry breaking, supersymmetry or extra dimensions. Latest experimental data from the Atlas \cite{Atlas} and CMS \cite{CMS} experiments put stringent limits and constraints on the SUSY parameter space \cite{PDG}.  Moreover there is no sign of new particles that should accompany dynamical symmetry breaking or extra dimensions.  A possibility not often explored is however  that the electroweak symmetry sector and breaking  is due to a combination of both dynamical mechanism and supersymmetry (see for example \cite{Bardeen1}-\cite{Bardeen2}).

 The best sources of information about possible dynamical mechanisms  \cite{Bardeen3}, \cite{Cvetic} stem from the study of low energy QCD \cite {Jora1}-\cite{Jora7} or superQCD \cite{Seiberg1}-\cite{Seiberg4} with the associated low energy effective models like Nambu Jona Lasinio models or linear sigma models. A main ingredient in constructing low energy effective models of QCD, since  fundamental knowledge of the intrinsic dynamics is often  absent, is to rely on symmetries like chiral symmetry and the anomaly structure of the theory, especially the $U(1)_A$ and trace anomalies. In \cite{Jora1}-\cite{Jora7} we proposed and developed a generalized linear sigma model with two chiral nonets that can describe with very good results the low mass scalar and pseudoscalar spectrum and some of the associated phenomenology.
 In these models adequate terms to mock up the anomalies are introduced based on the algebra of currents associated to a particular symmetry. It turns out that in order to exactly satisfy the anomalies it is necessary to introduce logarithmic terms dependent on the matter fields. For example the gluon axial anomaly requires the presence in the Lagrangian of terms such as:
 \begin{eqnarray}
 bF^{a\mu\nu}\tilde{F}^a_{\mu\nu}\ln[\frac{\det M}{\det M^{\dagger}}],
 \label{axterre}
 \end{eqnarray}
 where $F^{a\mu\nu}$ is the gluon tensor, $\tilde{F}^a_{\mu\nu}$ is the dual tensor and $M$ is the chiral nonet of scalar and pseudoscalar mesons made of a quark and an antiquark. In Eq. (\ref{axterre}) the gluon field can be integrated out leaving purely logarithm terms in the Lagrangian.
This kind of set-up which is specific to the construction of linear sigma models is the source of inspiration for the work presented here. However instead of coupling the logarithmic terms with some field in the Lagrangian we will introduce pure logarithmic operators. The final term will be the difference of two logarithmic operators dependent on the fermion degrees of freedom.  Due to the anticommutativity of the fermion variables, in the absence of any vacuum condensates this operator is designed to vanish.  In the presence of vacuum condensates however things change radically since from the start the main structure of the operator may not respect the electroweak symmetry or supersymmetry for example. The connection with the low energy effective model and with the dynamical mechanism of the Nambu Jona Lasinio type is straightforward.  We then apply the main electroweak experimental results concerning the Higgs mass, vacuum expectation value and couplings with the fermions to determine the masses and condensates in this approach of the main supersymmetric particles. All our calculations are approximate and the operators we introduce are by no means unique but only minimal.

To conclude we will start with a Lagrangian that is symmetric under the full electroweak group $U(1)_Y \times SU(2)_L \times SU(3)_c$ and even supersymmetric but contains no elementary scalars and end up with a scenario in which both electroweak symmetry and supersymmetry are broken dynamically by introducing an effective operator that it is in certain limit completely irrelevant.

In section II we present our operator, the interactions that it introduces and also the low energy constraints that must be satisfied. Section III contains an evaluation  of the various vacuum condensates and masses. Section IV is dedicated to discussions and conclusions.

\section{Set-up}

We consider the standard model or the supersymmetric standard model with all the fields and interactions but in the absence of the Higgs doublet and the electroweak breaking. Higher dimension operators can be also included in this set-up. We introduce the operator:
\begin{eqnarray}
T_1=2N_s\frac{\Lambda^4}{\pi^4}\ln[{\rm Tr}[\pm\frac{\bar{q}_Ld_R\bar{d_R}q_L}{\Lambda^6}+...]],
\label{firstop}
\end{eqnarray}
where the dots stand for all other gauge invariants formed out of the same quark or lepton doublets and singlets and also possible supersymmetric four fermion terms constructed from the same supersymmetric fermions. The second term that we introduce is:
\begin{eqnarray}
T_2=-N_s\frac{\Lambda^4}{\pi^4}\ln[{\rm Tr}[\pm\frac{(\bar{q}_Ld_R\bar{d_R}q_L)^2}{\Lambda^{12}}+...]],
\label{res774665}
\end{eqnarray}
where the dots stand for the same terms as in $T_1$, squared. Here $\Lambda$ is the cut-off of the theory to be determined later and $N_s$ some arbitrary constant.

This is by no mean an unique choice but only one of the minimal ones. In both $T_1$ and $T_2$ the terms may appear positive or negative but the overall coefficient is $\pm1$. Before going further we note that one can raise the arguments of the logarithms at any power without problems with the corresponding scaling of the coefficient in front. Then one can raise the argument of the logarithm in $T_1$ at power $N_1/4$ where $N_1$ corresponds to the total number of fermion degrees of freedom and $T_2$ to $\frac{N_1}{8}$ to notice that the argument of the logarithm will contain apart from a constant in front for both $T_1$ and $T_2$ a product of all fermion degrees of freedom. Here we took into account the anticommutativity of the fermion variables. Since the coefficients in front the logarithms are fixed such that
$2N_s\frac{4}{N_1}=N_s\frac{8}{N_1}$  it is obvious that the general field contribution of the operators $T_1$ and $T_2$ is zero and that they cancel up to a constant contribution. As such these operators are completely irrelevant.
Nevertheless we shall introduce this operators that do not modify essentially the initial Lagrangian and assume that at some scale one of the fermion fields form a two fermion condensate due to some strong dynamics. Note that this approach is inspired by the axial term in the effective linear sigma model in low energy QCD \cite{Jora1}.

The full terms of interest  is thus $T_1+T_2$ and we will be interested only in the associated vacuum structure. Note that the corresponding phenomenology could be quite rich but here we will be concerned only with some particular aspects.
From the point of view of the vacuum structure we will associate to each fermion-antifermion bilinear that may condensate the notation $S_i$, where $i$ is the fermion species (Note that this is just a useful notation and $S_i$ does not necessarily represent a scalar bound state and has mass dimension $3$). The terms are scaled such that to lead to the following normalization:
\begin{eqnarray}
T_1+T_2=2N_s\frac{\Lambda^4}{\pi^4}\ln[\sum_i\frac{(S_i+\alpha_i)^2}{\Lambda^6}+....]-
N_s\frac{\Lambda^4}{\pi^4}\ln[\sum_i\frac{(S_i+\alpha_i)^4}{\Lambda^{12}}+...].
\label{vac35442}
\end{eqnarray}
Here $\alpha_i$ is the vacuum condensate for the fermion $i$. Then the logarithms in Eq. (\ref{vac35442}) may be expanded around the vacuum condensates which leads to:
\begin{eqnarray}
&&T_1+T_2=4N_s\frac{\Lambda^4}{\pi^4}\frac{1}{\sum_k\alpha_k^2}\Bigg[\alpha_i-\alpha_i^3\frac{\sum_k\alpha_k^2}{\sum_k\alpha_k^4}\Bigg]S_i+
\nonumber\\
&&2N_s\frac{\Lambda^4}{\pi^4}\frac{1}{\sum_k\alpha_k^2}\Bigg[1-\frac{2\alpha_i^2}{\sum_k\alpha_k^2}-3\alpha_i^2\frac{\sum_k\alpha_k^2}{\sum_k\alpha_k^4}+4\alpha_i^6\frac{\sum_k\alpha_k^2}{(\sum_k\alpha_k^4)^2}\Bigg]S_i^2+
\nonumber\\
&&4N_s\frac{\Lambda^4}{\pi^4}\frac{1}{\sum_k\alpha_k^2}\sum_{i\neq j}\Bigg[-2\frac{\alpha_i\alpha_j}{\sum_k\alpha_k^2}+4\alpha_i^3\alpha_j^3\frac{\sum_k\alpha_k^2}{(\sum_k\alpha_k^4)^2}\Bigg]S_iS_j+....
\label{finalres735546663}
\end{eqnarray}
Note that in $T_1$ and $T_2$ we have the freedom  to change the signs of the various fermion terms in the arguments of the logarithms but not the scale. We will use this feature later in our calculations.

We make the notations:
\begin{eqnarray}
&&G=4N_s\frac{\Lambda^4}{\pi^4}\frac{1}{\sum_i\alpha_i^2}
\nonumber\\
&&\sum \pm\alpha_i^2=\alpha_0^2
\nonumber\\
&&x_i=\frac{\alpha_i}{\alpha_0}
\nonumber\\
&&\sum_i\pm x_i^4=s.
\label{notte646}
\end{eqnarray}
Then one can infer:
\begin{eqnarray}
\sum_i\pm x_i^2=1.
\label{constr663554}
\end{eqnarray}
In what follows we will determine the parameters $x_i$ such that to fit the main low energy information that we have about the standard model. First we note the main relation in this approach:
\begin{eqnarray}
G\alpha_i[1\pm \frac{x_i^2}{s}]=\pm m_i,
\label{res663455}
\end{eqnarray}
where generically the vacuum expectation value is obtained through a gap equation as in \cite{Bardeen3}.
Then for the quarks for example we have:
\begin{eqnarray}
Gm_i(1\pm\frac{x_i^2}{s})\frac{3}{4\pi^2}\Lambda^2=\pm m_i
\label{res663553}
\end{eqnarray}
Since we expect the top quark to have the largest condensate we make from the beginning the choice (which is actually  a result of low energy phenomenology data):
\begin{eqnarray}
1-\frac{t^2}{s}=-1,
\label{firstconstr}
\end{eqnarray}
where $t=\frac{\alpha_t}{\alpha_0}$ and $\alpha_t$ is the top quark vacuum condensate.
Then in order to get plausible values for the other quarks and for the gap equation to still hold we must have $x_i\ll1$ for all other quarks.
From Eq. (\ref{res663553}) we infer:
\begin{eqnarray}
G=\frac{4\pi^2}{3\Lambda^2}.
\label{gres6455}
\end{eqnarray}
Our first hypothesis is that the top bound  quark states are assimilated with the Higgs and are the main responsible for the electroweak symmetry breaking.
For that let us consider how the Higgs coupling may arrive. We follow \cite{Bardeen3} to introduce the effective Lagrangian:
\begin{eqnarray}
{\cal L}=y_i'\frac{1}{\sqrt{2}}\bar{\Psi}_i\Psi_ih-\frac{m_0^2}{2}h^2,
\label{eff34422}
\end{eqnarray}
where $h$ is an auxiliary field. We solve the equation of motion for the auxiliary field to determine:
\begin{eqnarray}
h=\sum_iy_i'\frac{1}{\sqrt{2}m_0^2}\bar{\Psi}_i\Psi_i.
\label{res663442}
\end{eqnarray}
We introduce the expression in Eq. (\ref{res663442}) into Eq. (\ref{eff34422}) to obtain:
\begin{eqnarray}
&&{\cal L}=\frac{1}{4m_0^2}\sum_{i,j}y_i'y_j'\bar{\Psi}_i\Psi_i\bar{\Psi}_j\Psi_j\approx
\nonumber\\
&&\frac{1}{4m_0^2}y_t^{\prime2}\bar{\Psi}_t\Psi_t\bar{\Psi}_t\Psi_t-\frac{1}{2m_0^2}\sum_{i\neq t}y_t'y_i'\bar{\Psi}_t\Psi_t\bar{\Psi}_i\Psi_i,
\label{final45665}
\end{eqnarray}
where the subscript $t$ refers to the top quark and $i$ to all the others. Here we took into account the fact that the Yukawa coupling is much larger for the top quark than for the rest of the quarks.
Next we need to compare the results in Eq. (\ref{final45665}) to those in Eq. (\ref{finalres735546663}) to see if they make sense in our set-up. We obtain:
\begin{eqnarray}
\frac{G}{2}\Bigg[1-2t^2-3\frac{t^2}{s}+4\frac{t^6}{s^2}\Bigg]=\frac{1}{4m_0^2}y_t^{\prime2}.
\label{first4553441}
\end{eqnarray}
Eq. (\ref{first4553441}) however needs a small corrections. In order to generate a correct term we need to subtract form $\alpha_0^2$ in $G$ the vacuum condensate for the top quarks and then reconsider it.
This leads to:
\begin{eqnarray}
\frac{G'}{2}\Bigg[1-2t^2-3\frac{t^2}{s}+4\frac{t^6}{s^2}\Bigg]=\frac{1}{4m_0^2}y_t^{\prime2},
\label{first4553441}
\end{eqnarray}
where $G'=G\frac{1}{1-t^2}$.  Then Eq. (\ref{first4553441}) becomes:
\begin{eqnarray}
\frac{G}{2}\Bigg[1-2t^2-3\frac{t^2}{s}+4\frac{t^6}{s^2}\Bigg]=\frac{1}{4m_0^2}y_t^{\prime2}(1-t^2),
\label{first4553441}
\end{eqnarray}
In the quark loop approximation \cite{Bardeen3}, \cite{Cvetic} the heavy quark degrees of freedom are integrated out in the energy interval $[\mu,\Lambda]$ which leads  to the approximate effective Lagrangian:
\begin{eqnarray}
{\cal L}=Z_h\partial^{\mu}h\partial_{\mu}h +y_i'\frac{1}{\sqrt{2}}\bar{\Psi}_i\Psi_ih-\frac{m_0^2}{2}h^2,
\label{lagrsec345552}
\end{eqnarray}
where only the terms of interest are retained. The field $h$ is rescaled by $h'=Z_h^{1/2}h$ and the low energy conditions of electroweak symmetry breaking lead to:
\begin{eqnarray}
\frac{1}{\sqrt{2}}\frac{y_t'}{\sqrt{Z_h}}\langle h'\rangle=y_t\frac{1}{\sqrt{2}}v.
\label{res442332}
\end{eqnarray}
where $v=\langle h'\rangle$ is known electroweak vacuum $v\approx 246.22$ GeV. One  defines the renormalized top quark coupling constant as:
\begin{eqnarray}
y_t=\frac{y_t'}{\sqrt{Z_h}}.
\label{couplt77676}
\end{eqnarray}
We now need to compare the second term on the second line of Eq. (\ref{final45665}) with our set-up in Eq. (\ref{finalres735546663}). This leads to the constraint:
\begin{eqnarray}
2tx_iy_t^{\prime 2}(1-t^2)\Bigg[1-2t^2-2\frac{t^2}{s}+4\frac{t^6}{s^2}\Bigg]^{-1}=y_t'y_i'.
\label{form64775665}
\end{eqnarray}
Noting that also $y_i=\frac{y_i'}{\sqrt{Z_h}}$ we further obtain:
\begin{eqnarray}
2tx_i y_t(1-t^2)\Bigg[1-2t^2-2\frac{t^2}{s}+4\frac{t^6}{s^2}\Bigg]^{-1}=y_i.
\label{finrels64788}
\end{eqnarray}
Using,
\begin{eqnarray}
\frac{y_t}{y_i}=\frac{m_t}{m_i}=\frac{\alpha_t}{\alpha_i}=\frac{t}{x_i},
\label{somrel7756664}
\end{eqnarray}
we arrive at the equality:
\begin{eqnarray}
2t^2a=2t^2(1-t^2)\Bigg[1-2t^2-2\frac{t^2}{s}+4\frac{t^6}{s^2}\Bigg]^{-1}\approx 1,
\label{constr775664}
\end{eqnarray}
where we denoted:
\begin{eqnarray}
a=(1-t^2)\Bigg[1-2t^2-2\frac{t^2}{s}+4\frac{t^6}{s^2}\Bigg]^{-1}
\label{not6577464}
\end{eqnarray}

Next one can write in the quark loop approximation \cite{Cvetic} for the mass of the Higgs boson:
\begin{eqnarray}
m^2(\mu^2)=\frac{1}{Z_h}[m_0^2-\frac{3}{8\pi^2}y_t^{\prime 2}(\Lambda^2-\mu^2)].
\label{mass647738}
\end{eqnarray}
Using from Eq. (\ref{first4553441}) $m_0^2=y_t^{\prime  2}\frac{3\Lambda^2}{8\pi^2}a$ and setting $\mu^2=m_h^2$ we obtain (Note that we consider this the mass of the Higgs boson in broken phase):
\begin{eqnarray}
m_h^2=\frac{3}{8\pi^2}\Lambda^2y_t^2[a-1]\frac{1}{1-y_t^2\frac{3}{8\pi^2}}.
\label{constsec348899}
\end{eqnarray}

Furthermore from Eqs. (\ref{res663442}) and (\ref{first4553441}) we determine:
\begin{eqnarray}
\langle hy_t'\frac{1}{\sqrt{2}}\rangle=\langle h'y_t\frac{1}{\sqrt{2}}\rangle=G\alpha_t\frac{1}{a},
\label{anotherconstr665}
\end{eqnarray}
and:
\begin{eqnarray}
v\approx\sqrt{2}m_t\frac{1}{a}.
\label{res6657759}
\end{eqnarray}
This reinforces our previous finding that $a\approx1$.

At this point we have the necessary constraints for the low energy set-up.

\section{Vacuum condensates }
We first solve the system of equations obtained from the constraints of section II:
\begin{eqnarray}
&&t^2=2s
\nonumber\\
&&t^2=\frac{1}{2a}
\nonumber\\
&&a=(1-t^2)\Bigg[1-2t^2-2\frac{t^2}{s}+4\frac{t^6}{s^2}\Bigg]^{-1}.
\label{res7746638}
\end{eqnarray}
which leads to one single set of acceptable solutions:
\begin{eqnarray}
&&t=0.6815
\nonumber\\
&&s=0.2322
\nonumber\\
&&a=1.0765.
\label{firstsetofls8}
\end{eqnarray}
With the values in Eq. (\ref{firstsetofls8}) we solve Eq. (\ref{constsec348899}) to determine $\Lambda=2275$ GeV.

We then note that the condition of $a\approx 1 $ is automatically fulfilled by the value in Eq. (\ref{firstsetofls8}). Next we observe that $t^2\neq1$ which means that the theory has other significant vacuum condensates.
Since we cannot associate these to the other fermions of the standard model we conclude that the theory must contain supersymmetric particles. We thus include in our initial set-up gauginos and  gravitinos (Note that we are interested only in fermion states).

First we shall determine the gap equation for the gravitino. We start with the gravitino propagator \cite{Weinberg} in the euclidean space:
\begin{eqnarray}
&&P^{\mu\nu}=(\eta^{\mu\nu}+\frac{q^{\mu}q^{\nu}}{mg^2})(-iq^{\rho}\gamma^{\rho}+mg)-
\nonumber\\
&&\frac{1}{3}(\gamma^{\mu}-i\frac{q^{\mu}}{m_g})(iq^{\rho}\gamma^{\rho}+m_g)(\gamma^{\nu}-i\frac{q^{\nu}}{m_g}).
\label{prop75888}
\end{eqnarray}
This leads for the gravitino condensate in the general approach introduced in section II:
\begin{eqnarray}
\alpha_g=\langle\frac{1}{2}\bar{\Psi}^{\mu}\Psi_{\mu}\rangle=\frac{1}{24\pi^2m_g}\Lambda^2[\Lambda^2+6m_g^2(1-\frac{m_g^2}{\Lambda^2}\ln[\frac{m_g^2+\Lambda^2}{m_g^2}])],
\label{gvarr900}
\end{eqnarray}
and to the gap equation:
\begin{eqnarray}
G\alpha_g(1+\frac{y^2}{s})=\pm m_g,
\label{gap466363}
\end{eqnarray}
where $y=\frac{\alpha_g}{\alpha_0}$.
Noting that,
\begin{eqnarray}
\frac{\alpha_g}{\alpha_t}=\frac{y}{t},
\label{res664553443}
\end{eqnarray}
and introducing this result in Eq. (\ref{gap466363}) we obtain:
\begin{eqnarray}
m_g=\pm m_t\frac{y}{t}[1+\frac{y^2}{s}].
\label{res7766455}
\end{eqnarray}
 Eqs. (\ref{gvarr900}) and (\ref{res664553443}) lead to the consistency condition:
\begin{eqnarray}
\frac{1}{24\pi^2m_g}\Lambda^2[\Lambda^2+6m_g^2(1-\frac{m_g^2}{\Lambda^2}\ln[\frac{m_g^2+\Lambda^2}{mg^2}])]=\frac{3}{4\pi^2}\Lambda^2m_t\frac{y}{t},
\label{gvarr900}
\end{eqnarray}
Substituting $m_g$ from Eq. (\ref{res7766455}) we solve for $y$ to obtain $y=1.291$. The corresponding mass of the gravitino is then $m_g=2681$ GeV which is very close and slightly larger than the cut-off scale. This result should come as no surprise since in general gap equations of the Nambu Jona Lasinio type lead naturally to masses of order $\Lambda$, the cut-off scale of the theory.

Next we need to consider gauginos. We start with the gaugino associated to the group $U(1)_Y$ and denote $u=\frac{\alpha_u}{\alpha_0}$ where $\alpha_u$ is the corresponding condensate. The gap equation reads:
\begin{eqnarray}
G\alpha_u[1\pm\frac{u^2}{s}]=\pm m_u
\label{gapu67766}
\end{eqnarray}
which further leads to the condition:
\begin{eqnarray}
\frac{4\pi^2}{3\Lambda^2}\frac{2}{16\pi^2}[\Lambda^2-m_u^2\ln[\frac{m_u^2+\Lambda^2}{m_u^2}]]=1.
\label{gapforu657748}
\end{eqnarray}
It turns out that Eq. (\ref{gapforu657748}) has no solution compatible with our initial set-up.  Consequently the mass of this gaugino is zero in first order and should be generated through loop corrections. Furthermore $u$ is either zero or satisfy the equation $u^2=\pm s $.

For the gauginos associated to the group $SU(2)_L$ the gap equation reads (Here $p=\frac{\alpha_p}{\alpha_0}$ where $\alpha_p$ is the corresponding vacuum condensate):
\begin{eqnarray}
G(1+\frac{p^2}{s})\alpha_p=\pm m_p.
\label{resp97758}
\end{eqnarray}
Using,
\begin{eqnarray}
\frac{\alpha_p}{\alpha_t}=\frac{p}{t},
\label{res553443}
\end{eqnarray}
we obtain:
\begin{eqnarray}
m_p=\pm m_t\frac{p}{t}[1+\frac{p^2}{t}].
\label{res664553443}
\end{eqnarray}
We introduce the result in Eq. (\ref{res664553443}) into Eq. (\ref{res553443}) to get:
\begin{eqnarray}
[1+\frac{p^2}{s}][1-\frac{m_p^2}{\Lambda^2}\ln[\frac{m_p^2+\Lambda^2}{m_p^2}]]-2=0,
\label{finaleq7736}
\end{eqnarray}
where we used:
\begin{eqnarray}
\alpha_p=\frac{6}{16\pi^2}\Lambda^2[1-\frac{m_p^2}{\Lambda^2}\ln[\frac{m_p^2+\Lambda^2}{m_p^2}]]m_p.
\label{res42443}
\end{eqnarray}
We solve Eq. (\ref{finaleq7736}) to obtain $p=1.2966$ and further $m_p=2713$ GeV.

Next step is to consider gluinos. We denote $x=\frac{\alpha_x}{\alpha_0}$ where $\alpha_x$ is the gluino vacuum condensate with:
\begin{eqnarray}
\alpha_x=\frac{16}{16\pi^2}\Lambda^2m_x,
\label{res7746}
\end{eqnarray}
The gap equation is:
\begin{eqnarray}G\alpha_x(1-\frac{x^2}{s})=\pm m_x,
\label{gl885776}
\end{eqnarray}
and we are looking for solution with relatively small $\frac{m_x}{\Lambda}$ knowing that larger values do not fit in our set-up.
The gap equation reduces to:
\begin{eqnarray}
1-\frac{x^2}{s}=\pm\frac{3}{4},
\label{res74635553}
\end{eqnarray}
with two solutions $x_1=0.2410$ corresponding to the plus sign on the right hand side of Eq. (\ref{res74635553}) and  $x_2=0.6375$ corresponding to the minus sign on the right hand side of Eq. (\ref{res74635553}). Using,
\begin{eqnarray}
\frac{\alpha_x}{\alpha_t}=\frac{4}{3}\frac{m_x}{m_t}=\frac{x}{t},
\label{res55467777}
\end{eqnarray}
we determine $m_x=\frac{3}{4}\frac{x}{t}m_t$ which leads to the two values: $m_{x1}=46$ GeV associated to $x_1$ and $m_{x2}=121.4$ GeV associated to $x_2$.
These values for the gluino masses are well below the PDG limits \cite{PDG} $m_g\gg 700-1870$ GeV. Then one needs to consider particular cases of supersymmetric models where these limits are circumvented (see for example \cite{Shirai}).

\section{Discussion and conclusions}
In section III we computed the relevant values for the quantities $x_i=\frac{\alpha_i}{\alpha_0}$.  These values must satisfy the constraints which were the starting point of the whole approach:
\begin{eqnarray}
&&\sum_i\pm x_i^2=1
\nonumber\\
&&\sum_i\pm x_i^4=s.
\label{constr775664}
\end{eqnarray}
Here it is important to notice two things. First the values $x_i$ for the rest of the fermions which were not discussed in section III are very small but their contributions can sum up to some  small, finite quantity.
The set-up might contain fermions like for example extra Majorana neutrinos or the $U(1)_Y$ gaugino that might have the parameter $x_i^2=s$. Considering all these facts it turns out that there are three scenarios that might work (Note that $s\approx0.2322$):

1) Scenario I. Here:
\begin{eqnarray}
&&t^2-y^2-x_2^2+p^2+4s\approx 1.00148
\nonumber\\
&&t^4+y^4-x_2^4-p^4+4s^2\approx0.21778.
\label{firstcsne554664}
\end{eqnarray}
In this case the gluino has mass $m_{x2}$ and there are four Majorana fermions with masses almost zero and with $x_i^2=s$.
\vspace{1cm}

2) Scenario II:
\begin{eqnarray}
&&t^2-y^2+x_1^2+p^2+2s\approx 1.00418
\nonumber\\
&&t^4+y^4+x_1^4-p^4+2s^2\approx 0.27845.
\label{scnen455667}
\end{eqnarray}
In this case gluino has mass $m_{x1}$ and there are two Majorana fermions with masses almost zero and $x_i^2=s$.
\vspace{1cm}

3) Scenario III:
\begin{eqnarray}
&&t^2-y^2+x_2^2+p^2+s\approx 1.11760
\nonumber\\
&&t^4+y^4+x_2^4-p^4-s^2\approx 0.27845.
\label{sceny7777}
\end{eqnarray}
For this case gluino has mass $m_{x2}$ and there is only one Majorana fermion with mass almost zero and  $x_i^2=s$.

In all these scenarios we took into account the relative signs of the terms in the operators $T_1$ and $T_2$ that correspond to the particular fermion contributions and solutions as computed in section III.

In this work we developed not a supersymmetric model but an effective low energy image of a possible supersymmetric theory consistent with the main low energy experimental data. We start from  the presumption of dynamical electroweak and supersymmetry breaking in the absence of any elementary Higgs boson and introduce a logarithmic  operator that it is irrelevant when symmetries are valid but may account in the low energy limit for a dynamical mechanism of symmetries breaking.   Our analysis is by no means exact or exhaustive but it leads to some low energy phenomenological answers that may reveal at least partially some hints with regard to the high energy behavior of the theory. We propose some values for the vacuum condensates and masses of some  of the supersymmetric fermions like gauginos and gravitinos but  any of the possible scenarios of UV completion associated to them is beyond the scope of our work.

\end{document}